# Entropy driven multi-photon frequency up-conversion

Assaf Manor[1], Nimrod Kruger[2], and Carmel Rotschild[1,2,3*]

*1. Russell Berrie Nanotechnology Institute, Technion − Israel Institute of Technology, Haifa 32000, Israel*
*2. Grand Technion Energy Program, Technion − Israel Institute of Technology, Haifa 32000, Israel*
*3. Department of Mechanical Engineering, Technion − Israel Institute of Technology, Haifa 32000, Israel*

*Correspondence to: carmelr@technion.ac.il*

Frequency up-conversion of few low-energy photons into a single high-energy photon, greatly contributes to imaging, light sources, detection and other fields of research[1-3]. However, it offers negligible efficiency when up-converting many photons. This is because coherent process are fundamentally limited due to momentum conservation requirements[4], while in incoherent up-conversion the finite intermediate states lifetime requires huge intensities. Thermodynamically, conventional incoherent up-conversion is driven by the internal energy of the incoming photons. However, a system can also drive work through change in its collective properties such as entropy. **Here we experimentally demonstrate entropy driven ten-fold up-conversion from 10.6μ to 1μm at internal efficiency above 27% and total efficiency above 10%.** In addition, the emitted radiance at 1μm exceeds the maximal possible Black-Body radiance of our device, indicating emitter's effective-temperature that is considerably above the bulk-temperature. This work opens the way for up-conversion of thermal-radiation, and high-temperature chemistry done at room-temperature.

Traditional frequency up-conversion effects include coherent (second, third and parametric up-conversion[2,5,6] and incoherent (two photon[7–9] and multi-photon absorption[10]) processes, yet they all offer negligible efficiency when converting many (~Ten) photons into a single high-energy photon. This negligible conversion-efficiency is due to the large momentum mismatch between the pump and the produced frequencies, and the need for many photons to interact simultaneously requires huge intensities. **For these reasons the record efficiency of tenfold up-conversion is lower than 0.01%, and was achieved under pulse excitation at intensities of $10^{15}$ W/cm; many orders of magnitude higher than currently available continuous wave [CW] sources**[11].

To find efficient up-conversion mechanisms for partially-incoherent light, we begin by relating the work involved in the process of photoluminescence to the work performed by a thermodynamic system. Here, the potential energy of each photon is defined by its chemical potential, $\mu$ [12, 13]. In this view, absorption of energetic photon followed by emission of a red shifted photon while releasing heat to the environment acts as an "optical heat pump" [14], driven by the difference in the chemical potentials of the photons. Photoluminescence efficiency can reach unity due to the reduction in photon energy defining its spontaneously. Can we have similar spontaneous up-conversion process? In the complete thermodynamic picture a spontaneous process is accompanied by a reduction in its Gibbs free energy, $G$ [15]:

$dG=d[PV-TS+\sum \mu N]$, which may occur even if the photon internal energy increases. Here *P, V, T* and *S* are collective properties of the system: *P* is pressure, *V* is volume, *T* is temperature, *S* is the entropy which is proportional to the number of populated states in the system, and $\sum \mu N$ is the sum of the chemicals potentials.

In most optical systems *P* and *V* are constant, thus the change in *PV* can be excluded from generating work. An exception to this general rule is the phenomenon of Sonoluminescence, where UV emission is generated as sound is converted into light through a drastic change in *PV*[16].

To the best of our knowledge, the change in the temperature and entropy of a system, *TS,* has never been used in frequency up-conversion. Here, we do just that: as we show below, we utilize the change in *TS* for efficient tenfold frequency up-conversion.

For lasers, entropy relates to the spatial and temporal coherences. As the entropy is lower, the laser coherence is higher and the Gibbs free energy is maximal, allowing maximal light concentration. Such a radiation source has a high "Brightness-temperature", which is defined as radiance in a specific wavelength that is equal to a Black-Body radiance at the same temperature and the same wavelength[17-18]. The term radiance refers to the amount of radiation power per wavelength per solid-angel per area. Such a high brightness-temperature radiation, if absorbed under perfect conditions, can equate the targets temperature to the brightness-temperature. A naive implementation of this concept may be the use a low photon energy laser (such as $CO_2$ laser with photon energy of 0.1eV) to increase a body's temperature, thereby enhancing its Black-Body radiation of energetic photons. However, although this concept can be highly efficient up-conversion mechanism, it is challenging to realise simply because there are no materials that remain stable at such high temperatures. Also, the broad thermal emission is less attractive for many applications, such as communication and detection where the information is frequency-dependent. An ideal device should operate at room temperature and emit spectrally narrow up-converted light.

Our approach is to use low photon energy but high brightness-temperature source, such as a laser, to excite only specific modes to a high effective-temperature, while keeping the bulk at relatively low temperature. Although under such non-equilibrium conditions the term temperature is not well defined, "Effective-temperature" describes a population of a specific phonon-mode that corresponds to Boltzmann's thermal population at that temperature. An emitter that is efficiently coupled to these highly excited modes, shares their effective-temperature, and emits a spectrally narrow radiation at high brightness-temperature, considerably higher than the bulk temperature. Thus far, utilizing such ideas for frequency conversion have never been explored, even though evidence for high effective-temperature was detected in semiconductors long ago[19]. Those pioneering experiments revealed that optical phonons could be excited to an effective-temperature of *3700K* while the bulk (GaAs crystal) remained at *290K*. Figure 1 shows the free-energy reduction accompanied by up-conversion, as conventionally sketched for the photon internal-energy flow in photoluminescence (down-conversion).

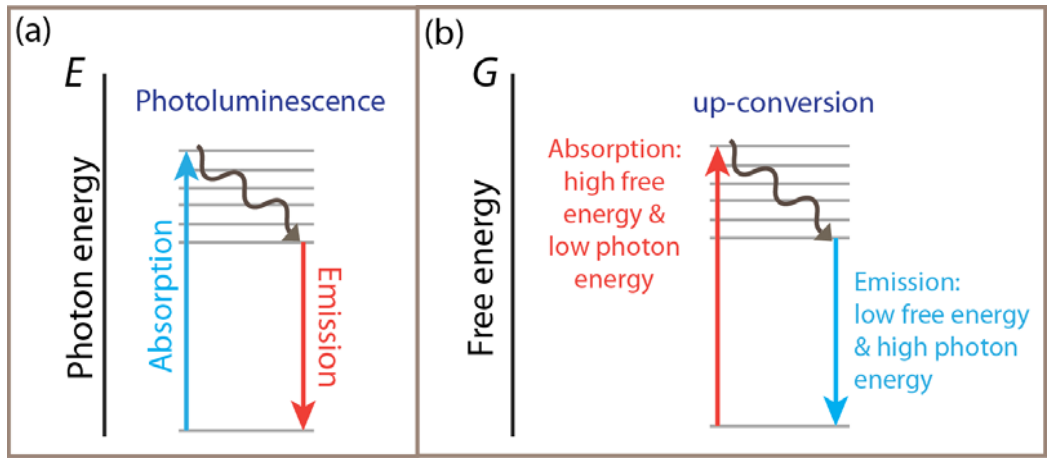

**Fig. 1: Photoluminescence (a) and Free-energy diagram for frequency up-conversion process (b)**

To experimentally demonstrate this concept, we produce $Yb_2O_3$ doped $SiO_2$ spherical cavities at diameters between *30μ* and *300μ*. We focus a CW $CO_2$ laser beam at *10.6μ* wavelength, and power levels lower than *1W*, onto the spherical resonator within an integrating sphere. The resonator emission between 0.4μ and 11μ wavelength is detected by a spectrograph and calibrated against a Black-Body source at 1200C (see Methods Section). Figure 2 shows the experimental setup (Fig. 2a), as well as a picture of the spherical cavity (Fig. 2b). Such a cavity, when un-doped, is known to support whispering-gallery modes [WGMs] with Q-factors exceeding $10^7$ [20]. Its first mode is simulated in Fig. 2c using finite element method simulation[21].

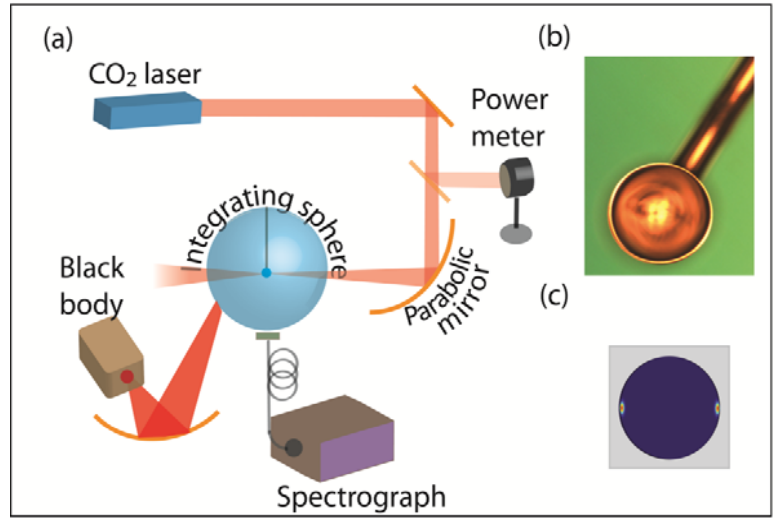

**Fig. 2:** The experimental setup (a) the $SiO_2$ resonator (b) and its first WGM (c)

We begin by measuring the resonator's radiance at the near-infrared [NIR] under excitation of 684mW. We first capture the emission image using a Si CCD camera, (detecting wavelengths shorter than 1.1μ) in order to gain information about the spatial distribution of the radiation, which is essential for calculating the radiance. Next we measure the power spectrum at the NIR region using the calibrated integrating sphere. Normalizing the power spectrum with the emission area gives a minimal value for isotropic radiance. We compare the measured radiance with the Black-Body and $Yb_2O_3$ radiances at 1650C[22, 23], which is the melting point of $SiO_2$, hence serving as the upper limit for the device stability.

**Results:**

Figure 3 summarises the typical results of many experiments done with over 20 different resonators. Figures 3a and 3b show the captured emission up to 1.1μm wavelength. The emission appears as the spatial distribution of WGMs[24] with the highest Q-factor. In Fig. 3a the emission corresponds to the first mode, similar to the simulation results shown in Fig. 2c. Figure 3b shows the emission from the first few modes. Figure 3c compares the resonator (of Fig. 3b) radiance with the radiance of bulk $Yb_2O_3$ at 1650C. As can be seen, the resonator emission has a sharp peak at 980nm (black solid line), which can be attributed to Ytterbium emission. This sharp peak is very different than the emission of bulk $Yb_2O_3$ under thermodynamic equilibrium, where most of its energy is in the thermal region of the spectrum, at wavelengths longer than 1μ (blue dotted line in Fig. 3c). Actually, the emission resembles photoluminescence of Ytterbium pumped by energetic photons. Figure 3d shows emission measurements in the IR-region, showing residual radiation beyond 2.4μ wavelength. Comparing these results with Black-Body radiance at 1650C reveals **that the 980nm peak radiance is four times that of a Black-Body radiance at the maximum possible device temperature** (black and dotted red lines at 3d).

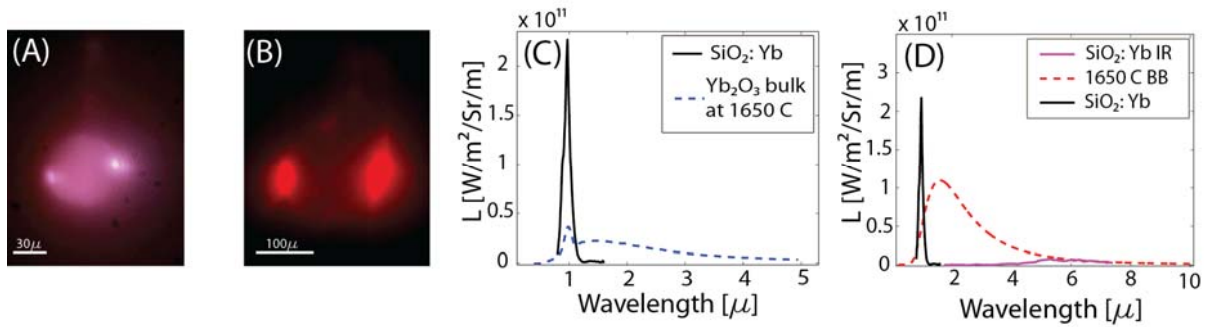

**Fig. 3:** CCD imaging of a single (a) and multimode emission (b). Device radiance versus $Yb_2O_3$ radiance at 1650C, (c) Device and Black-Body at 1650C radiances (d).

Furthermore, **the emission at 980nm contains 27% of the total emission power, and 1% of the total input power deposited into the system.** Such efficiency is many orders of magnitude better than any related prior art we are aware of. In fact, we are able to observe more than an order of magnitude enhancement in Ytterbium emission under partial pressure of 300mbar, measured through the vacuum chamber window. This indicates a **total conversion efficiency in excess of 10%.** This can be attributed to the reduction in the surface thermal losses.

**Discussion**:

The high efficiency and the narrow spectral emission can be explained by the emitter's high brightness-temperature, much above bulk temperature. There are three major evidences supporting this claim: **i.** At the infra-red wavelength ($5\mu$-$10\mu$) where silica is highly emissive, its thermal emission fits a Black-Body radiation nearly at melting temperature, 1650C. This evidence reflects the true temperature of the resonator near thermodynamic equilibrium, also evident by the stability of the device, which doesn't melt. **ii.** $Yb_2O_3$ emission at 1650C should have a thermal lobe at wavelengths longer than 1μm. This doesn't appear in our experiment. Instead, the emission exhibits a sharp single peak. **iii.** The radiance of the $Yb_2O_3$ is four times the Black Body radiance at 1650C, which is the maximal possible radiance under thermodynamic equilibrium. This is true even when taking into account the radiation enhancement by the Purcell's effect of the high-Q WGM. Based on these three observations we argue that **$Yb_2O_3$ is at considerably higher Brightness-temperature than the bulk**.

To conclude, we presented a new frequency up-conversion mechanism and experimentally demonstrated efficient tenfold frequency up-conversion driven by entropy. This may open new directions in various fields of research. In solar energy for example, our concept opens efficient ways for up-converting sub-bandgap solar photons to wavelengths where photovoltaics are most efficient. In another example, the possibility to excite modes to a very high effective-temperature while the bulk remains at low temperature opens the way for "hot chemistry" experiments carried out at room temperature. Hot chemistry may include conversion of sunlight into stored chemical energy, by employing a high temperature endothermic thermo-chemical response such as the MgO-Mg redox, which exhibits energy recovery above 42% [25]. Each of these examples may revolutionize our ability to harvest the sun's energy for humanity needs.

**Methods:**

1. **Sample preparation:**

    Commercial 125μ fibers (Thorlabs) were positioned in the focus of a 20W $CO_2$ laser (Synrad). The fiber tip was melted and fed into the focal point until the formation of a sphere, easily seen under the microscope. The Sphere was subsequently dip-coated in a methanol:$Yb_2O_3$ nano-crystals suspension (1ml methanol:100mg $Yb_2O_3$ ), and then melted again by a short laser expose of about half a second to gain a smooth surface sphere. For smaller spheres fabrication, the fibers were etched in an HF solution until the desired diameter was reached, prior to the melting process.

2. **Optical measurements:**

    The sample's radiance was measured by placing it in a white integrating sphere (LabSphere) for the NIR range and in a self-designed gold coated sphere for the IR range. A CO2 laser (Access lasers) at various intensities with stability of 2% was focused onto the sample by a gold parabolic mirror and ZnSe lens.

    In addition, a 1200$C$ calibrated black-body source (CI-systems) was focused inside the sphere for calibration purposes. The luminescence signals were chopped by an optical chopper (Stanford Research Systems) and amplified by a lock-in amplifier (Stanford Research Systems), after passing through a spectrograph equipped with the appropriate gratings for the different spectral ranges (Oriel Instruments). In the NIR region the signals were measured by Ge detector (Judson Technologies), and InGaAs camera

(Andor Technology). In the IR region between 2µm and 10µm, the signal was detected using InSb and MCT (InfraRed Associates) detectors .

For the measurements under vacuum, the samples were put in a dewar with optical windows connected to a vacuum system. The laser was focused onto the sample through a ZnSe window. The luminescence signal was measured by imaging the sample through the dewar window into the spectrograph equipped with a spectroscopic InGaAs camera.

### 3. Recipe for entropy driven up-conversion:

To our view, there are three major conditions to reach entropy driven up-conversion. **i.** The low internal photon energy pump source must be at high brightness-temperature. In our experiment this condition is easily matched with the $CO_2$ laser, operating at a single (or few) mode, at a brightness-temperature in the order of $10^{10}\,C$. **ii.** The excitation rate to a set of modes must exceed their dumping rate to other modes; In our experiment this condition is met by the high absorption cross-section of $SiO_2$, which absorbs the 10.6µ $CO_2$ laser radiation within a sub-wavelength depth on the surface through its vibronic-states[1,2]. **iii** The emitter must be efficiently coupled to the vibronic-states. This condition is satisfied in $Yb_2O_3$ and other rare-earths, as have been shown in Refs. 3, 4.

**Methods references:**

1. A. D. McLachlan, F. P. Meyer, Temperature dependence of the extinction coefficient of fused silica for CO2 laser wavelengths, *Appl. Opt.* **26**, 1728–1731 (1987).

**Acknowledgments:** This report was partially supported by the by the Russell Berrie Nanotechnology Institute (RBNI), and the Grand Technion Energy Program (GTEP) and is part of The Leona M. and Harry B. Helmsley Charitable Trust reports on Alternative Energy series of the Technion and the Weizmann Institute of Science. We also would like to acknowledge partial support by the Focal Technology Area on Nanophotonics for Detection.

**Author Contributions:**

http://www.nature.com/authors/policies/authorship.html

All authors contributed extensively to the work presented in this paper.

**Figure legends:**

**Fig. 1**: Photoluminescence (a) and Free-energy diagram for frequency up-conversion process (b)

**Fig. 2:** The experimental setup (a) the $SiO_2$ resonator (b) and its first WGM (c)

**Fig. 3:** CCD imaging of a single (a) and multimode emission (b). Device radiance versus $Yb_2O_3$ radiance at 1650C, (c) Device and Black-Body at 1650C radiances (d).